\documentclass[runningheads]{llncs}
\usepackage[T1]{fontenc}
\usepackage{newtxtext}       %
\usepackage[varvw]{newtxmath}       
\usepackage{booktabs}
\usepackage{subcaption}
\usepackage{multirow}
\usepackage{graphicx}
\begin{document}
\title{Partially-Averaged Navier-Stokes simulation of turbulent annular Poiseulle flow}
%
%
\author{Li-Toong Yap\inst{1,2} \and
Marten Klein\inst{1,2} \and
Heiko Schmidt\inst{1,2}
}


%
\authorrunning{L-T Yap et al.}
%
\institute{Lehrstuhl Numerische Strömungs und Gasdynamik, Brandenburgische Technische Universität Cottbus-Senftenberg, Siemens-Halske-Ring 15A, 03046 Cottbus, Germany\and
Scientific Computing (SC) Lab, Energie-Innovationszentrum (EIZ), Brandenburgische Technische Universität Cottbus-Senftenberg, Siemens-Halske-Ring 15A, 03046 Cottbus, Germany  
}

\maketitle              
\begin{abstract}
Turbulent annular Poiseulle flow is commonly encountered in various engineering applications, such as heat exchangers. The effects of the inner wall transverse curvature on the flowfield, characterised by the ratio of inner and outer radii $\eta = R_{i}/R_{o}$, have been studied previously using Direct Numerical Simulations (DNS). 
Scale-resolving simulations based on a Reynolds-Averaged Navier–Stokes (RANS) framework offer a potentially cheaper alternative to traditional Large Eddy Simulations (LES); however it is not clear if such models are able to capture the turbulence anisotropy in the turbulent annular pipe flow. 
In the present study, turbulent annular Poiseulle flow at $Re_{D_{h}} = 8900$ is performed with the Partially-Averaged Navier Stokes (PANS) model \cite{girimaji_partially-averaged_2006}, a variable-resolution turbulence model that is based on the conventional RANS modelling. The ability of the PANS model to capture the effects of the transverse wall curvature is found to be comparable with DNS and other RANS-based scale-resolving approaches \cite{jakirlic_curvature_2024} despite its relative simplicity and reduced computational cost.

\end{abstract}
\section{Introduction}
Turbulent annular Poiseulle flow occurs in various engineering applications, such as heat exchangers and electro-hydrodynamics. The dynamics of the momentum transport are affected by the difference in transverse curvature between the two cylinders, which can be characterised in terms of the radius ratio $\eta=R_{i}/R_{o}$, where $R_{i}$ and $R_{o}$ are the inner and outer radius respectively. The most noticeable effect of the inner wall transverse curvature on the momentum transport is the asymmetry of the flowfields in the gap. The annular Poiseulle flow is a generalisation of the plane channel and circular pipe flows where the former is achieved for $\eta\rightarrow 1$ while the latter is for $\eta\rightarrow0$ (although the presence of the ever small inner cylinder prevents it from becoming a circular pipe flow). The effects of the transverse curvature have been studied previously in Direct Numerical Simulations (DNS) \cite{bagheri_effects_2020,chung_direct_2002}. However, such studies can be computationally expensive. In recent years there have been interest in scale-resolving methods, different from traditional Large Eddy Simulations (LES), that are based on Reynolds Averaged Navier Stokes (RANS) models \cite{girimaji_partially-averaged_2006,jakirlic_extending_2015,menter_scale-adaptive_2010}. Of particular interest in this study is the Partially-Averaged Navier Stokes (PANS) methodology, first introduced in \cite{girimaji_partially-averaged_2006}. Most of its applications over the past few years have mainly been to external flows, with internal flows not investigated as much. The PANS methodology typically uses a Boussinessq-type closure for the unresolved components of the Reynolds stress tensor. Hence, the ability of such a closure to capture the transverse curvature effects is assessed in this study.
\section{Methodology}
\subsection{Governing equations}
In this study, the flowfield is treated as incompressible, neglecting the effect of buoyancy forces. 
Application of an arbitrary filter $\langle\cdot\rangle$ that is constant preserving and commutes with both time and spatial derivatives \cite{girimaji_partially-averaged_2006} to the instantaneous incompressible Navier-Stokes (NS) equations gives
\begin{equation}
\label{eqn: filtered NS equations}
\frac{\partial U_{i}}{\partial x_{i}} = 0,\quad
\frac{\partial U_{i}}{\partial t} +  U_{j}\dfrac{\partial  U_{i}}{\partial x_{j}}
   = -\dfrac{1}{\rho}\dfrac{\partial \tilde p }{\partial x_{i}} + \nu\dfrac{\partial^{2} U_{i}}{\partial x_{j}\partial x_{j}}- \dfrac{\partial\tau(V_{i}, V_{j})}{\partial x_{j}}, 
\end{equation}
where $U = \langle V\rangle$; $\tilde p = \langle p\rangle$. $V$ and $p$ represent the instantaneous velocity and pressure respectively. The filtering operation produces an additional term, the sub-filter stress (SFS) which is given by $\tau(V_{i}, V_{j}) = \langle V_{i}V_{j}\rangle - \langle V_{i}\rangle\langle V_{j}\rangle$ \cite{germano_turbulence_1992}. 
\subsection{Partially-Averaged Navier Stokes}
In the PANS framework, the SFS is closed similarly to the RANS framework. For this study, the Bousinessq hypothesis is applied to the SFS such that
\begin{equation}
     \tau(V_{i}, V_{j}) = -2\nu_{u}S_{ij} + \frac{2}{3}k_{u}\delta_{ij},
\end{equation}
where $S_{ij}$ is the resolved strain rate tensor, $k_u$ is the unresolved turbulent kinetic energy and $\nu_{u}$ is the unresolved eddy viscosity. 
Calculation of $\nu_{u}$ is done using the PANS $k$-$\omega$ Shear Stress Transport (SST) model \cite{lakshmipathy_assessment_2011}, which is based on the RANS $k$-$\omega$ SST model of Menter \cite{menter2003}. The PANS transport equations are given as 
 \begin{equation}
    \label{eqn: PANS k omega sst equations}
    \begin{aligned}
        \frac{\partial k_{u}}{\partial t} +  U_{j}\dfrac{\partial k_{u}}{\partial x_{j}}&=
        \widetilde{P_{ku}} - \beta^{*}\omega_{u} k_{u}  + \frac{\partial}{\partial x_{j}}\left[\left(\nu + \nu_{u}\sigma_{k}\frac{f_{\omega}}{f_{k}}\right)\frac{\partial k_{u}}{\partial x_{j}} \right]\\
        \frac{\partial \omega_{u}}{\partial t}+  U_{j}\dfrac{\partial \omega_{u}}{\partial x_{j}}&=
        \frac{\gamma}{\nu_u} \widetilde{P_{ku}}
        -\left(\gamma\beta^{*}-\frac{\gamma\beta^{*}}{f_{\omega}}-\frac{\beta}{f_\omega}\right)\omega^{2}_{u}\\
        &+ \frac{\partial}{\partial x_{j}}\left[\left(\nu + \nu_{u}\sigma_{\omega}\frac{f_{\omega}}{f_{k}}\right)\frac{\partial\omega_{u}}{\partial x_{j}}\right] \\
        &+ 2(1-F_{1})\sigma_{\omega}\frac{f_{\omega}}{f_{k}}\frac{1}{\omega_{u}}\frac{\partial k_{u}}{\partial x_{j}} \frac{\partial \omega_{u}}{\partial x_{j}} \\
        \widetilde{P_{ku}} = &\min(P_{ku}, 10\beta^{*}k_{u}\omega_{u}),\;\;P_{ku} = -\tau(V_{i}, V_{j})\frac{\partial U_{i}}{\partial x_{j}}
    \end{aligned}
\end{equation}
The unresolved eddy viscosity, the ratio of unresolved to total turbulent kinetic energy and the ratio of unresolved to total specific dissipation rate are given as
\begin{equation}
\label{eqn: eddy viscosity expression}
    \nu_{u} = \min\left(\frac{k_u}{\omega_u}, \frac{a_1 k_u}{SF_2}\right), \;\;\; f_k = \frac{k_u}{k}, \;\;\; f_\omega = \frac{\omega_u}{\omega}
\end{equation}
where $S$ is the invariant measure of the resolved strain-rate tensor. The parameters $f_k$ and $f_\omega$ represent the physical resolution parameters of the PANS model. Setting $f_k = 0$ represents the DNS solution as $k_u=0$, while $f_k=1$ represents the results from a RANS simulation since there is no resolved turbulent kinetic energy. $\omega$ is tied to the dissipation rate as $\epsilon = \beta^{*}k\omega$. The ratio of the unresolved to total dissipation is
\begin{equation}
    f_{\epsilon} = \frac{\epsilon_u}{\epsilon}
\end{equation}
As such, $f_\omega$ can be related to $f_\epsilon$ via the relation
\begin{equation}
\label{eqn: fomega relation}
    f_\omega = \dfrac{\omega_u}{\omega} = \dfrac{\dfrac{\epsilon_u}{\beta^{*}k_u}}{\dfrac{\epsilon}{\beta^{*}k}} = \frac{f_{\epsilon}}{f_{k}}
\end{equation}

The remaining coefficients used in Equations \ref{eqn: PANS k omega sst equations} and \ref{eqn: eddy viscosity expression} are similar to the RANS counterpart and can be found in \cite{menter2003}. The auxiliary functions $F_1$ and $F_2$ are given by 
\begin{equation}
\begin{aligned}
    &F_1 = \tanh \left\{ \min \left[ \max \left( \frac{\sqrt{k_u}}{\beta^* \omega_u y}, \frac{500 \nu}{y^2 \omega_u} \right), \frac{4  \alpha_{\omega 2u} k_u}{CD_{k\omega} y^2} \right]^4 \right\} \\
    &CD_{k\omega} = \max \left( \frac{2  \alpha_{\omega 2u}}{\omega_u} \frac{\partial k_u}{\partial x_j} \frac{\partial \omega_u}{\partial x_j}, 10^{-10} \right), \;\;\; \alpha_{\omega2u} = \alpha_{\omega2}\frac{f_\omega}{f_k}
\end{aligned}
\end{equation}
\begin{equation}
    F_2 = \tanh \left[ \max \left( 2 \frac{\sqrt{k_u}}{\beta^* \omega_u y}, \frac{500 \nu}{y^2 \omega_u} \right)^2 \right]
\end{equation}
where $y$ is the distance to the nearest wall.

A major choice in PANS simulations is the specification of the filter parameters $f_k$ and $f_\omega$. 
As most of the dissipation is assumed to occur at the small scales for high $Re$, $f_{\epsilon} = 1$ in this study as suggested by \cite{girimaji_partially-averaged_2006} , and $f_{\omega}$ is calculated based on Equation \ref{eqn: fomega relation}. 

\subsection{Computational set-up}
\begin{figure}[t]
    \centering
    \includegraphics[width=0.45\linewidth]{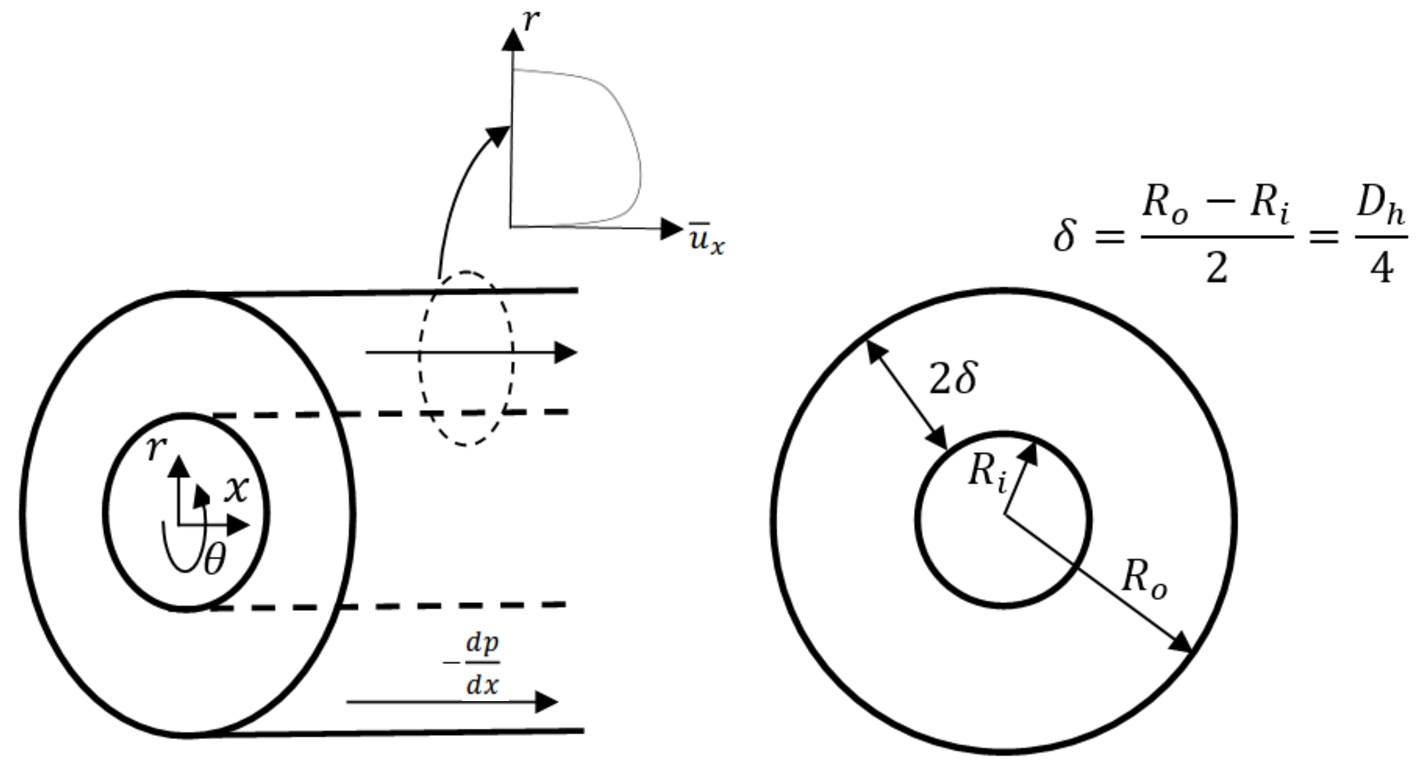}
    \caption{Sketch of the annular pipe configuration showing the driving mean pressure gradient and a qualitative mean velocity profile of the fully developed flow (\textbf{left}) as well as some relevant geometry parameters (\textbf{right}). }
    \label{fig:annular pipe schematic}
\end{figure}
The PANS equations were implemented in OpenFOAM \cite{weller_tensorial_1998} (v.9), an open-source CFD code based on the cell-centered finite-volume method. The pressure-velocity coupling is solved using the PIMPLE methodology, which is a combination of the pressure-implicit with splitting of operators \cite{issa_solution_1986} (PISO) methodology and the semi-implicit method for pressure linked equation \cite{patankar_numerical_2018} (SIMPLE) methodologies. The fully-developed turbulent flow is driven by a constant pressure gradient corresponding to the hydraulic Reynolds number $Re_{D_{h}} = U_{b}D_{h}/\nu =8900$, where $D_{h}=4\delta$, $\delta = 0.5(R_{o}-R_{i})$ and $U_{b}$ is the bulk velocity.
Periodic boundary conditions (BCs) were imposed at the inlet and outlet for all quantities. At the walls, no-slip BCs are specified for the velocity field, and zero gradient BCs are imposed for the pressure. The turbulence variables are set as \cite{Wilcox}
\begin{equation}
    k_{u} = 0, \;\;\; \omega_{u} = f_{\omega}\frac{6\nu}{\beta_{1}(\Delta y_{1})^{2}}
\end{equation}
where $\beta_{1} =0.075$. The BC for $k_{u}$ is enforced at the wall, while $\omega_{u}$ is enforced at the first cell from the wall with distance between the cell centre and wall $\Delta y_1$. Due to the resolution requirements required for the manifestation of the Tollmien-Schlichting wave instability, only $f_{k}$ values $\leq 0.20$ are considered \cite{klapwijk_evaluation_2020}. 

A structured grid was generated using \textit{blockMesh} for both radius ratios, with grading applied to the cells in the wall-normal direction such that the dimensionless distance from the inner or outer wall $y_{i/o}^{+} = |R_{i/o}-r|u_{\tau, i/o}/\nu$ was less than one. Table \ref{tab:mesh_details} summarises the geometrical and mesh characteristics used in this study, where $N_{x}$, $N_{r}$ and $N_{\theta}$ represent the number of cells in the axial, radial and azimuthal directions. Wall distances are non-dimensionalised using the respective friction velocities $u_{\tau, i/o}$ and kinematic viscosity $\nu$, with the exception of $\Delta x^{+}$ which uses $u_{\tau,o}$. The mesh used in this study is of similar size and resolution to the one used in \cite{jakirlic_curvature_2024}, where a sensitised RANS based on a Reynolds Stress Model (RSM) was used.

\begin{table}[t!]\centering
\caption{Mesh and geometry details}
\label{tab:mesh_details}
\begin{tabular}{@{}lcc@{}}
\toprule
($N_x$, $N_r$, $N_\theta$)   & \multicolumn{2}{c}{(150, 50, 360)}           \\
($L_x$, $L_r$, $L_\theta$)   & \multicolumn{2}{c}{(20$\delta$, 2$\delta$, $2\pi$)} \\ 
$Re_{D_{h}}$ & \multicolumn{2}{c}{8900} \\
\midrule
$\eta = R_{i}/R_{o}$              & 0.1                 & 0.5                \\ \midrule
$\Delta y_{i}^+$       & 0.6                 & 0.6                \\
$\Delta y_{o}^+$     & 0.58                & 0.58               \\
$(R_i \Delta \theta)^{+}$ & 0.694               & 5.06               \\
$(R_o \Delta\theta)^{+}$ & 5.62                & 10.12              \\
$\Delta x^{+}$          & 19.3                & 19.3               \\ \bottomrule
\end{tabular}
\end{table}
The convective and diffusive terms in the momentum transport equations were discretised using the central differencing scheme, while the convective terms in the transport equations for the turbulence quantities were discretised using a combination of central (90\%) and upwind differencing. The temporal discretisation was done using the second-order backward differencing scheme, with the time-step size automatically adjusted during run-time to ensure a maximum CFL number of 0.7. The steady state was assessed via the wall-shear stress on both walls, after which statistics were collected over approximately 20 flow-through times (FTT).

\section{Results}
\subsection{Velocity field}
\begin{figure}[t!]
    \centering
    \begin{subfigure}[b]{0.48\textwidth}
        \centering
        \includegraphics[width=\textwidth]{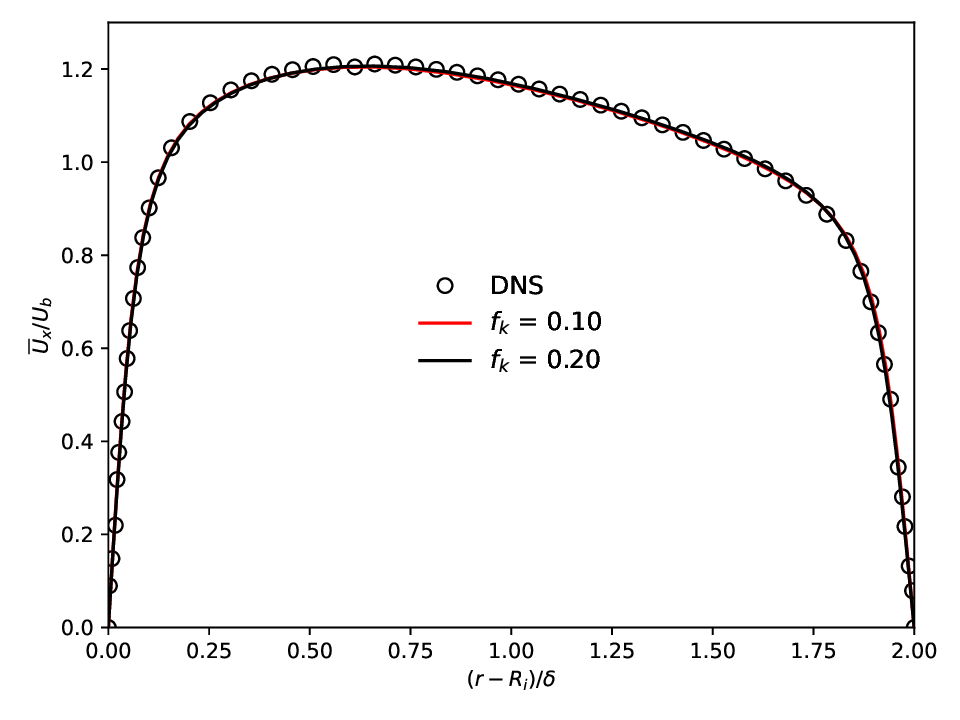} 
        \caption{$\eta=0.1$}
        \label{fig: UMean eta01}
    \end{subfigure}
        \hfill 
    \begin{subfigure}[b]{0.48\textwidth}
        \centering
        \includegraphics[width=\textwidth]{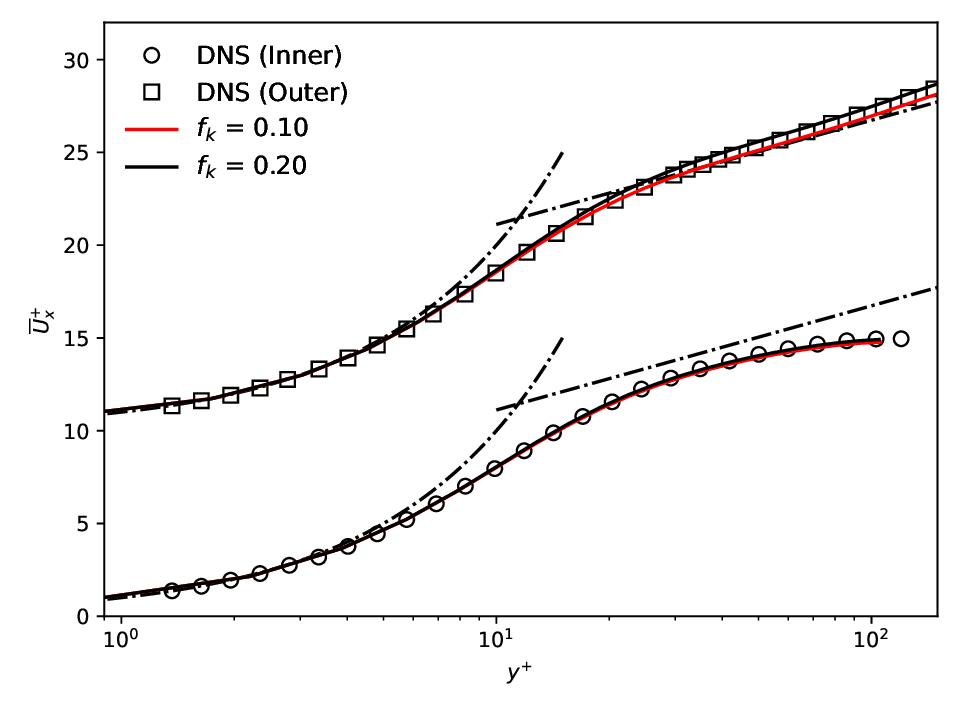} 
        \caption{$\eta=0.1$}
        \label{fig: UMean eta01 wall plot}
    \end{subfigure}
     \begin{subfigure}[b]{0.48\textwidth}
        \centering
        \includegraphics[width=\textwidth]{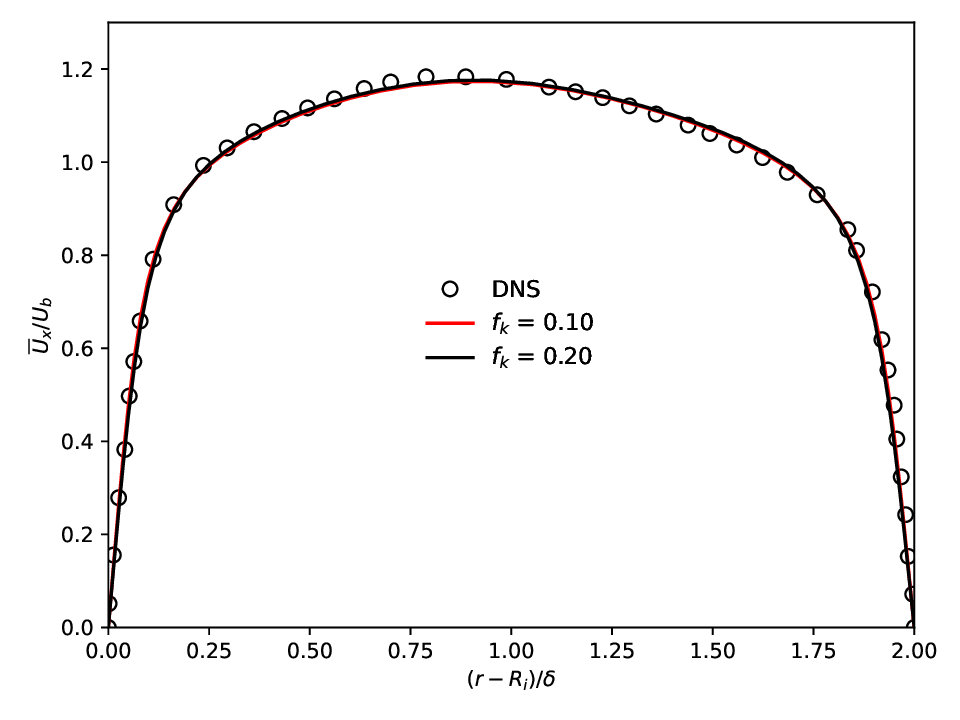} 
        \caption{$\eta=0.5$}
        \label{fig: UMean eta05}
    \end{subfigure}
        \hfill 
    \begin{subfigure}[b]{0.48\textwidth}
        \centering
        \includegraphics[width=\textwidth]{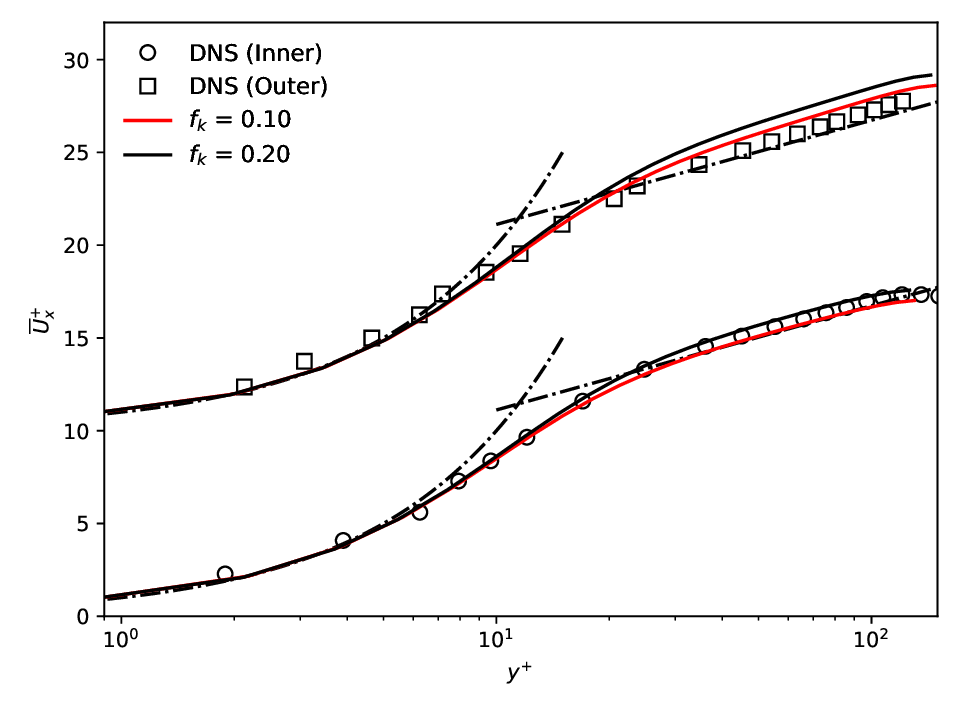} 
        \caption{$\eta=0.5$}
        \label{fig: UMean eta05 wall plot}
    \end{subfigure}
    \caption{Mean streamwise velocity profiles in global coordinates for each $\eta$, together with their respectively wall plots. Outer wall plots offset by 10 units upwards. Black dash-dotted lines represent the classical law of the wall. Reference DNS data taken from \cite{chung_direct_2002}.}
    \label{fig: Mean velocity profiles}
\end{figure}

\begin{table}[t!]
  \centering
  \setlength{\tabcolsep}{5pt}
  \caption{Bulk quantities obtained for the cases investigated, using different $f_k$.}
  \label{tab: bulk quantities for momentum}
  \begin{tabular}{@{}llcccccccc@{}}
    \toprule
    $Re_{D_h}$ & $\eta$ & $f_k$ & $\delta_{\mathrm{max}}/\delta$  & $\delta_{t,i}/\delta$ & $\delta_{t,o}/\delta$ & $C_{f, i}$ & $C_{f, o}$ & $Re_{\tau,i}$ & $Re_{\tau,o}$ \\
    \midrule
    \multirow{2}{*}{8900} & \multirow{2}{*}{0.1} 
      & 0.10 & 0.634 & 0.624 & 1.376 & 0.0133 & 0.00837 & 113.30 & 198.25 \\
      & & 0.20 & 0.644 & 0.634 & 1.366 & 0.0131 & 0.008 & 114.22 & 192.17 \\
    \midrule
    \multirow{2}{*}{8900} & \multirow{2}{*}{0.5} 
      & 0.10 & 0.913 & 0.914 & 1.086 & 0.00949 & 0.00795 & 140.28 & 152.58 \\
      & & 0.20 & 0.909 & 0.912 & 1.088 & 0.00894 & 0.00753 & 135.79 & 148.74 \\
    \bottomrule
  \end{tabular}
\end{table}
\begin{table}[t]
\centering
\setlength{\tabcolsep}{5pt}
\caption{DNS results from Chung \cite{chung_direct_2002} and Bagheri \cite{bagheri_effects_2020}, with the sensitised RSM results of Wang \cite{jakirlic_curvature_2024}}
\label{tab:dns results}
\begin{tabular}{@{}lcccccc@{}}
\toprule
       & \multicolumn{3}{c}{$\eta = 0.1$} & \multicolumn{3}{c}{$\eta=0.5$} \\  \midrule
   & Chung         & Bagheri   & Wang      & Chung          & Bagheri    & Wang      \\  \midrule
$C_{f,i}$    &   0.01300     &   0.0129 & 0.0131 &   0.00941                    &     0.00925    &   0.00928  \\
$C_{f,o}$    &  0.00810          &    0.00846  & 0.00817 &  0.00849                    &    0.00837     &  0.00811   \\
$\delta_{t,i}/\delta$ &   0.61         &   0.62 & 0.62 &    0.88                   &   0.88         & 0.87 \\
$\delta_{t,o}/\delta$ &   1.39        &   1.38   & 1.38&    1.12                   &     1.12       & 1.13 \\ \bottomrule
\end{tabular}
\end{table}
Figure \ref{fig: Mean velocity profiles} shows the mean velocity profiles for both radius ratios, plotted in global and wall coordinates. The mean velocity profiles in global coordinates show the increase in asymmetry, with the maximum location of the velocity $\delta_{max}$ shifting towards the inner wall, indicating that a thinner boundary layer forms over the inner wall. Furthermore, the asymmetry is accompanied by a notable deviation from the classical law-of-the-wall at the inner wall (Figure \ref{fig: UMean eta01 wall plot}), for which both $f_{k}$ values are able to predict. 
The normalised values of $\delta_{max}$ obtained are shown in Table \ref{tab: bulk quantities for momentum}. In comparison, values of $\delta_{max}/\delta =0.89$ for $\eta = 0.5$ and $0.64$ for $\eta=0.1$ were obtained in \cite{chung_direct_2002}, giving an error of approximately 2.2\% and 1.5\% respectively for $f_{k} = 0.10$.

\begin{figure}[t!]
    \centering
    \includegraphics[width=0.5\textwidth]{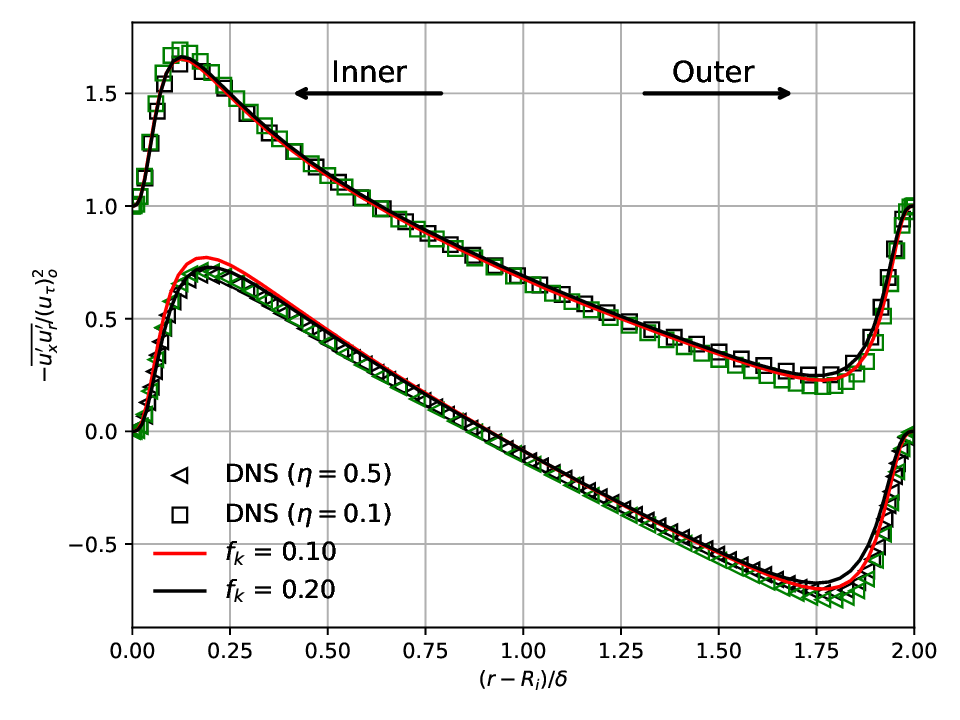}
    \caption{Reynold stress component $\overline{u^{'}_{x}u^{'}_{r}}$ in global coordinates, normalised by the friction velocity at the outer wall. Profile for $\eta=0.1$ is offset by 1 unit upwards. Black symbols represent DNS data taken from \cite{chung_direct_2002}, and green symbols represent DNS data from \cite{bagheri_effects_2020}.}
    \label{fig: Reynold stresses in global coordinates}
\end{figure}

The $Re_{\tau}$ values based on $u_{\tau, i/o}$ and local boundary layer thickness $\delta_{t,i/o}$ are also shown in Table \ref{tab: bulk quantities for momentum} together with the local skin friction coefficient $C_{f,i/o} = \tau_{w,i/o}/(0.5\rho U_{b}^{2})$. Also shown for convenience in comparison are $C_{f}$ and $\delta_{t}$ from the DNS \cite{bagheri_effects_2020,chung_direct_2002} and sensitised RSM \cite{jakirlic_curvature_2024}.
It is evident that decreasing $f_{k}$ values provide better agreement with the existing references. The largest deviations occur for the quantities at the outer wall for $\eta=0.5$, which is most likely due to the difference in the azimuthal resolution between both cases. 
Despite that, the results obtained for $\eta=0.5$ are in close agreement with the DNS data for $\overline{U_x}$ and $\overline{u^{\prime}_{x}u^{\prime}_{r}}$, as shown in Figures \ref{fig: Mean velocity profiles} and \ref{fig: Reynold stresses in global coordinates}.

\begin{figure}[t!]
    \centering
    \begin{subfigure}[b]{0.35\textwidth}
        \centering
        \includegraphics[width=\textwidth]{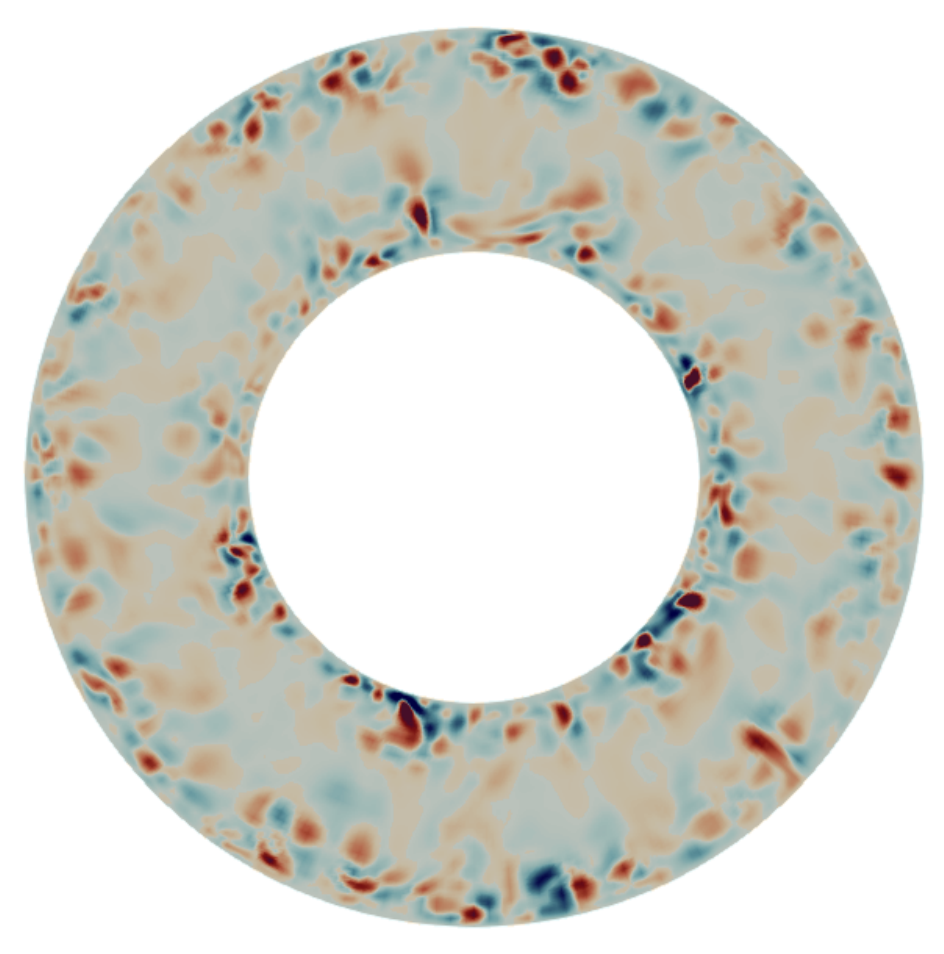} 
        \caption{$f_{k}=0.20$}
        \label{fig: Q-criterion-slice fk020}
    \end{subfigure}
    \begin{subfigure}[b]{0.42\textwidth}
        \centering
        \includegraphics[width=\textwidth]{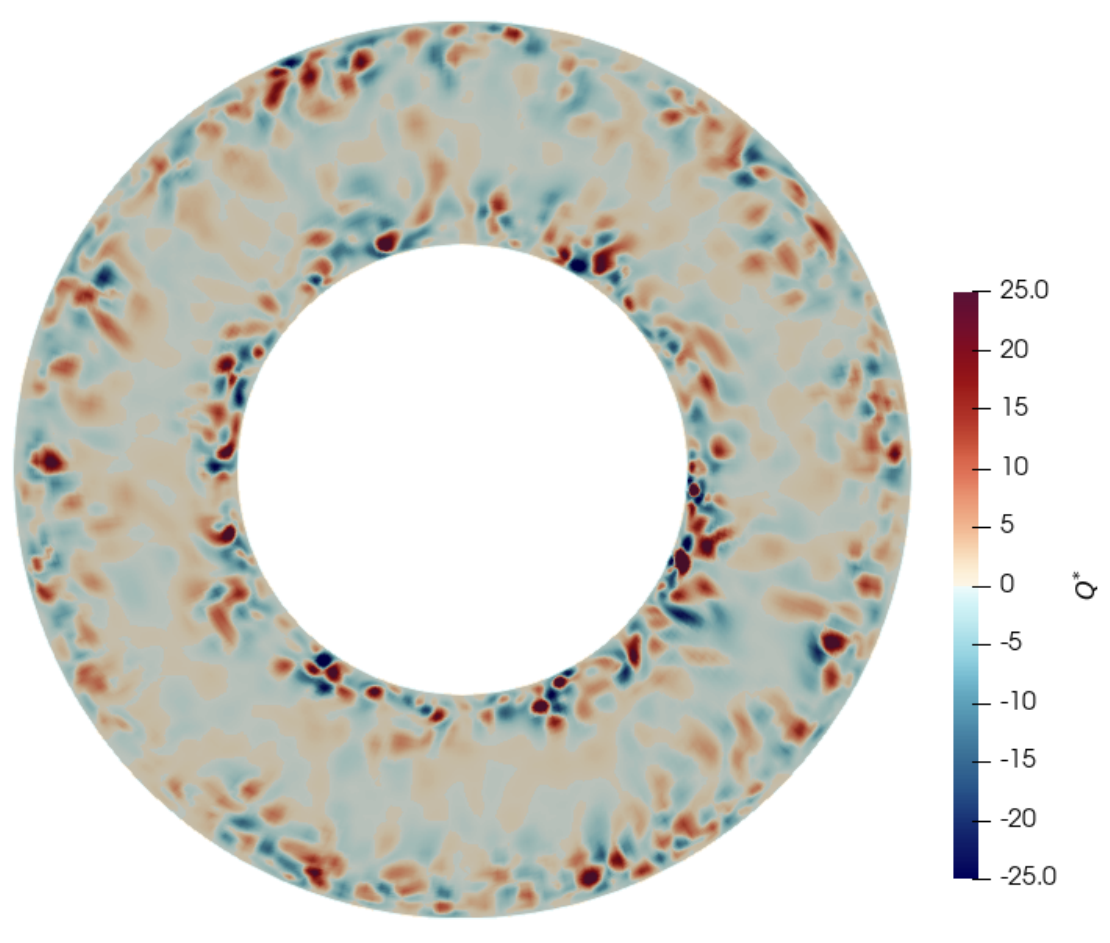} 
        \caption{$f_{k}=0.10$}
        \label{fig: Q-criterion-slice fk010}
    \end{subfigure}
    \caption{Slice of the normalised Q-criterion $Q^{*} = Q(D_{h}/U_{b})^{2}$ at $x$ = 10$\delta$.}
    \label{fig: Q-criterion slice}
\end{figure}  

The effect of $f_{k}$ on the turbulent structures is illustrated via the Q-criterion, defined as $Q = 1/2(|\Omega| - |S|)$ where $\Omega$ is defined as the resolved antisymmetric component of $\partial U_{i}/\partial x_{j}$ that represents local flow rotation. A decrease in $f_{k}$ results in more flow structures getting resolved as observed in Figures \ref{fig: Q-criterion slice} which shows the slice at $x = 10\delta$ of $Q$ for different $f_{k}$. There are more intense regions of vorticity (indicated by red) and shear (denoted by blue) at the inner and outer walls for the case of $f_{k} = 0.10$. 

\subsection{Turbulence Anisotropy}
In this section, as the results for $f_{k}= 0.20$ are found to be similar to $f_{k}=0.10$, only results for $f_{k}= 0.10$ are shown.
In order to quantify the effect of the transverse curvature on the turbulent structures, the turbulence anisotropy is investigated via the normalised deviatoric part of the Reynolds stress tensor: 
\begin{equation}
    b_{ij} = \frac{\overline{u^{\prime}_{i}u^{\prime}_{j}}}{\overline{u^{\prime}_{k}u^{\prime}_{k}}} - \frac{\delta_{ij}}{3}
\end{equation}
Furthermore, the second and third invariants of the normalised Reynolds stress tensor are
\begin{equation}
    II = \frac{1}{2}b_{ij}b_{ji}; \;\;\; III = -\frac{1}{3}b_{ij}b_{jk}b_{ki}
\end{equation}
Through these two invariants, the Lumley's two-component parameter F is defined as
\begin{equation}
    F = 1 - 9II + 27III
\end{equation}
\begin{figure}[p!]
    \centering
    \includegraphics[width=0.5\linewidth]{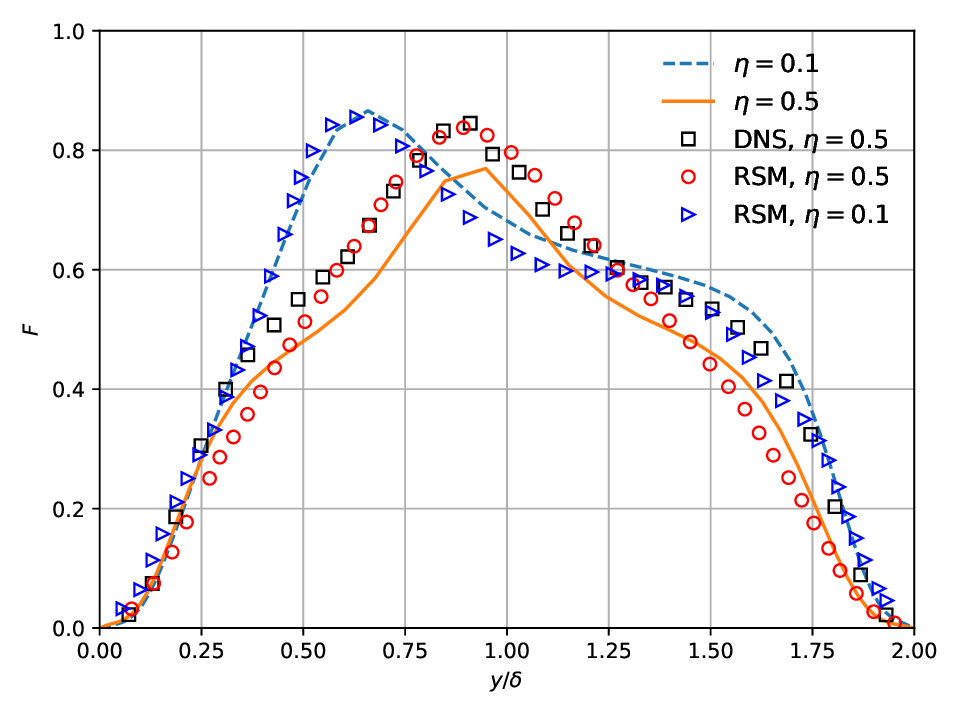}
    \caption{ $F$-component profile across the annular gap for both values of $\eta$, with $f_{k} = 0.10$. Reference DNS data taken from \cite{chung_direct_2002} and RSM data is from \cite{jakirlic_curvature_2024}.}
    \label{fig: F-component}
\end{figure}
\begin{figure}[p!]
    \centering
    \begin{subfigure}[b]{0.48\textwidth}
        \centering
        \includegraphics[width=\textwidth]{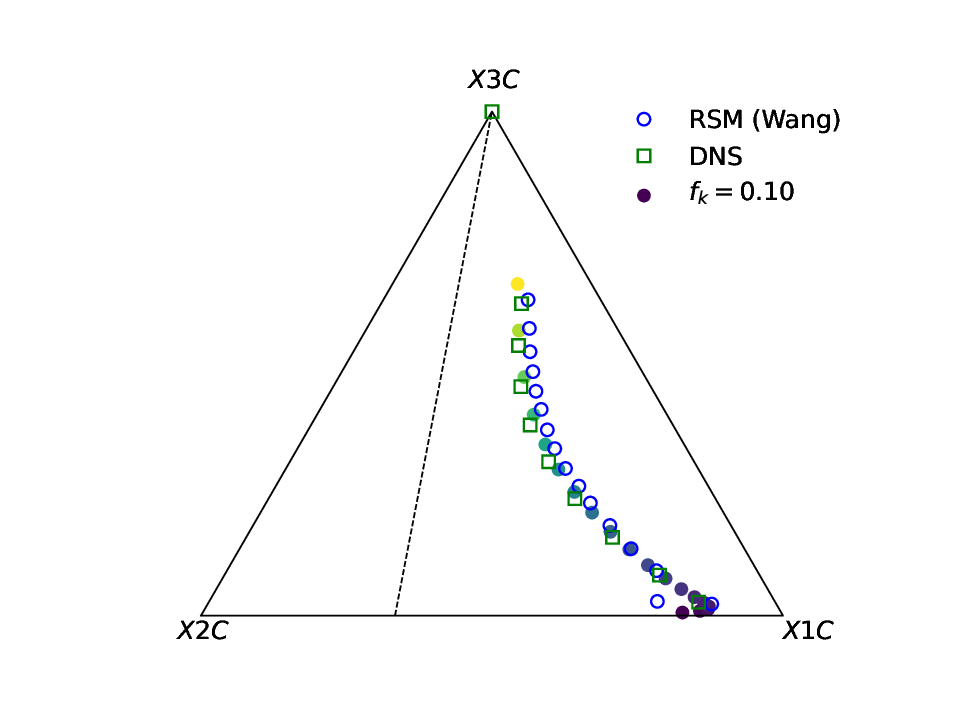} 
        \caption{Inner wall, $\eta=0.1$}
        \label{fig: barycentric plot inner eta01 fk010}
    \end{subfigure}
    \begin{subfigure}[b]{0.48\textwidth}
        \centering
        \includegraphics[width=\textwidth]{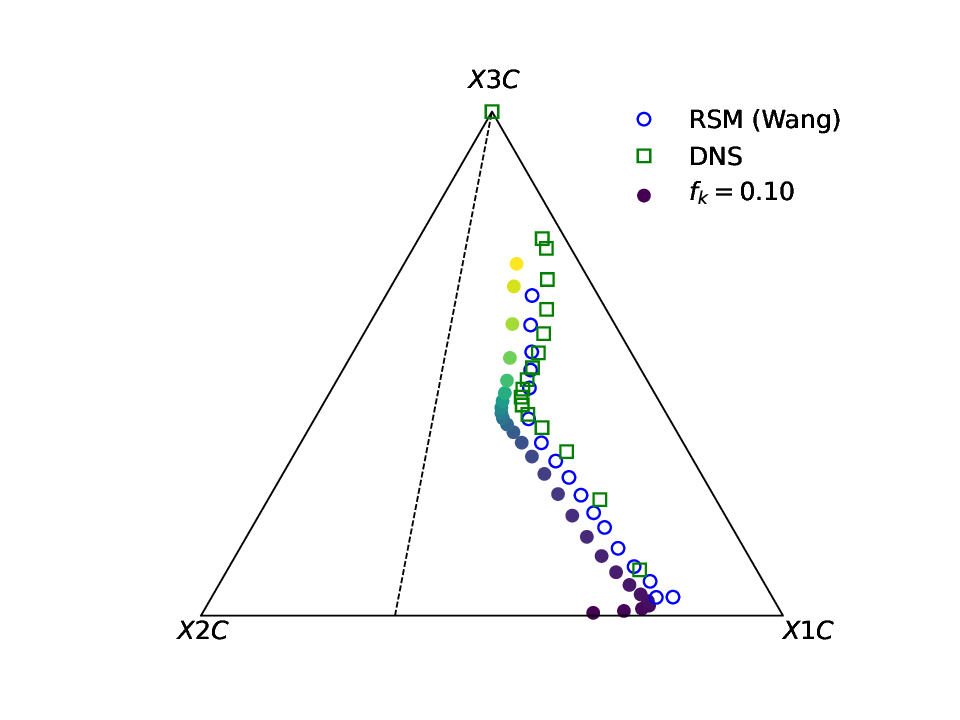} 
        \caption{Outer wall, $\eta=0.1$}
        \label{fig: barycentric plot outer eta01 fk010}
    \end{subfigure}
     \begin{subfigure}[b]{0.48\textwidth}
        \centering
        \includegraphics[width=\textwidth]{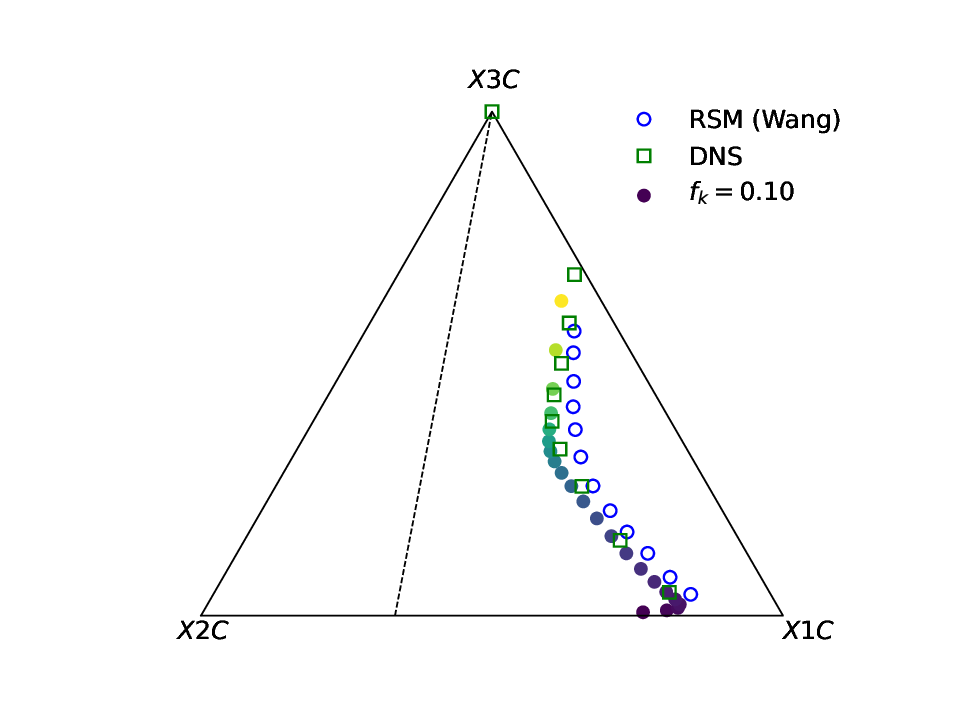} 
        \caption{Inner wall, $\eta=0.5$}
        \label{fig: barycentric plot inner eta05 fk010}
    \end{subfigure}
    \begin{subfigure}[b]{0.48\textwidth}
        \centering
        \includegraphics[width=\textwidth]{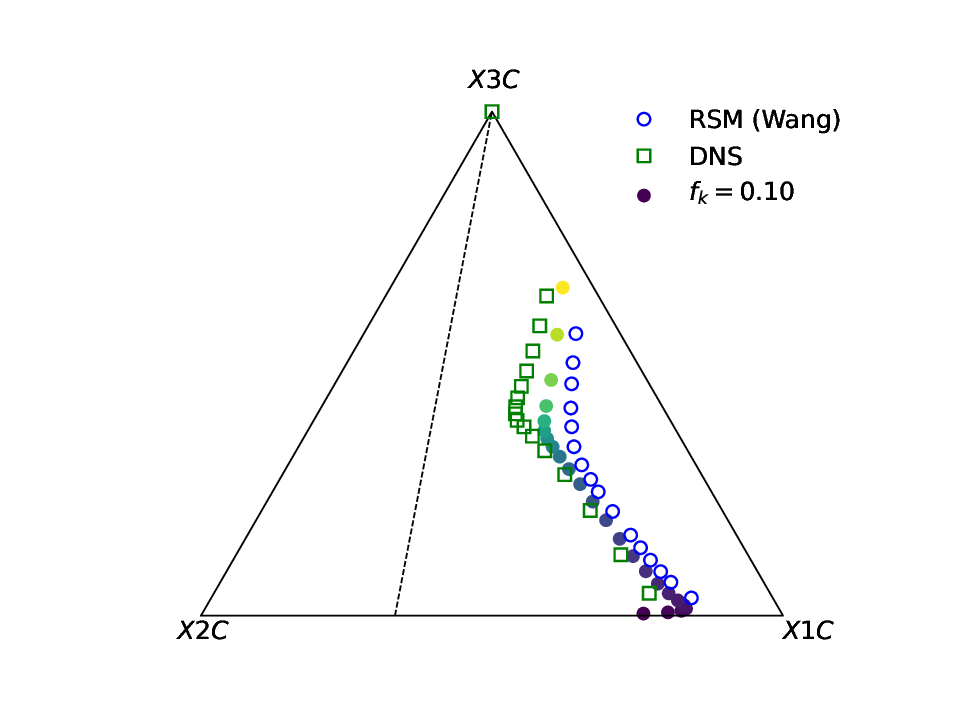} 
        \caption{Outer wall, $\eta=0.5$}
        \label{fig: barycentric plot outer eta05 fk010}
    \end{subfigure}
    \caption{Barycentric plots for $\eta = 0.1, 0.5$, $f_{k} = 0.10$. Colour encodes increasing $y^{+}$ from black to yellow. Reference DNS data taken from \cite{bagheri_effects_2020} and RSM data is from \cite{jakirlic_curvature_2024}. }
    \label{fig: barycentric plots}
\end{figure}  
where it takes a value of 0 for the turbulence state referred to as the two-component anisotropic turbulence and a value of 1 for a fully isotropic turbulence state. The profiles of the parameter F for $f_{k} = 0.10$ are shown in Figure \ref{fig: F-component} for the two radius ratios considered. For $\eta=0.5$, the profiles resemble that of the plane channel flow, although the maximum value of $F\approx0.79$ is under-predicted in comparison with that of the DNS reference \cite{chung_direct_2002} and the sensitised RANS-RSM reference \cite{jakirlic_curvature_2024} ($F\approx0.85)$. However, the profiles of the $F$-component from the sensitised RSM does not appear to have the same distinct bump in the vicinity of both walls that is present in the DNS and the current PANS results.

On the other hand, as $\eta$ decreases to 0.1, there is a noticeable shift of the peak $F$ towards the inner wall and increased values of $F$ in the boundary layer associated with the inner wall, indicating a lower amount of turbulence anisotropy. Unlike the $\eta=0.5$ case, the predicted maximum value of $F\approx0.84$ for $\eta=0.1$ is in close agreement with that of the sensitised RANS-RSM reference of $F\approx0.86$. The $F$-component profile is also in agreement with the reference RSM result.

In addition, the turbulence anisotropy can also be represented by a so-called barycentric map, a linear anisotropy invariant map as described by \cite{banerjee_presentation_2007}. This maps the profiles of the functional dependencies $f(II, III)$ to a set of barycentric coordinates $(x_{b}, y_{b})$ on a 2D equilateral triangle whose vertices represent the three limiting states of componentality. This mapping is shown in Figure \ref{fig: barycentric plots}. Noticeably, there is a mismatch away from the outer wall for $\eta = 0.5$ with the reference DNS and RSM data in Figure \ref{fig: barycentric plot outer eta05 fk010}. This mismatch is also reflected in the $F$-component profile for $\eta=0.5$ in Figure \ref{fig: F-component}. Despite the mismatch of the $F$-component near the outer wall, the profiles have a similar shape unlike the RSM data, which is also reflected in the barycentric map. Apart from that, the results are in good agreement with the reference data.  
\section{Conclusions}
In the present study, fully-developed turbulent annular pipe flow is simulated at $Re_{D_{h}} = 8900$ using the PANS k-$\omega$ SST model. The PANS model is able to reproduce the effects of the radius ratio $\eta$ on the 1st and 2nd order statistics using different values of $f_{k}$. For the case of $\eta=0.5$, it was found that there were some inaccuracies for the zero-crossing of the Reynolds stress component $\overline{u^{\prime}_{x}u^{\prime}_{r}}$ and the local skin friction coefficient $C_{f,i/o}$ in comparison with the reference DNS and sensitised RSM data, resulting in a mismatch of the $Re_{\tau, i/o}$. The ability of the PANS model to capture the turbulence anisotropy induced by the transverse curvature of the inner wall was also investigated using the $F$-component and the barycentric representation of the Lumley triangle. While the peak of the $F$-component profile for $\eta = 0.5$ was underpredicted by the PANS model, its shape was qualitatively similar in contrast with the sensitised RSM despite the latter predicting the peak values accurately. It is noted that the sensitised RSM study utilised similar resolutions and domains to the ones in this study. In addition, the trajectories on the barycentric map as predicted by the PANS model are in close agreement with the existing reference data. The largest discrepancies occurred at the outer wall of the $\eta=0.5$ case, although the barycentric coordinates predicted by the PANS model shared a similar shape with the DNS data as compared to the sensitised RSM.
The results obtained highlight the deficiencies in the various approaches for modelling the unresolved turbulent fluxes.

\begin{credits}
\subsubsection{\ackname} This research is supported by the German Federal Government, the Federal Ministry of Research, Technology and Space, and the State of Brandenburg within the framework of the joint project EIZ: Energy Innovation Center (project numbers 85056897 and 03SF0693A) with funds from the Structural Development Act for coal-mining regions. 

\end{credits}
%
%
%
\bibliographystyle{splncs04}
\bibliography{references}

@article{klapwijk_evaluation_2020,
    title = {Evaluation of scale-resolving simulations for a turbulent channel flow},
    volume = {209},
    doi = {10.1016/j.compfluid.2020.104636},   
    journal = {Computers \& Fluids},
    author = {Klapwijk, M. and Lloyd, T. and Vaz, G. and Van Terwisga, T.},
    year = {2020},
    pages = {104636}
}

@article{girimaji_partially-averaged_2006,
    title = {Partially-{Averaged} {Navier}-{Stokes} {Model} for {Turbulence}: {A} {Reynolds}-{Averaged} {Navier}-{Stokes} to {Direct} {Numerical} {Simulation} {Bridging} {Method}},
    volume = {73},
    issn = {0021-8936, 1528-9036},
    shorttitle = {Partially-{Averaged} {Navier}-{Stokes} {Model} for {Turbulence}},
    doi = {10.1115/1.2151207},
    number = {3},
    urldate = {2026-04-09},
    journal = {Journal of Applied Mechanics},
    author = {Girimaji, Sharath S.},
    month = may,
    year = {2006},
    pages = {413--421},
}

@inproceedings{lakshmipathy_assessment_2011,
    address = {Honolulu, Hawaii},
    title = {Assessment of alternative formulations for the specific-dissipation rate in {RANS} and variable-resolution turbulence models},
    isbn = {978-1-62410-148-9},
    doi = {10.2514/6.2011-3978},
    language = {en},
    urldate = {2026-04-09},
    booktitle = {20th {AIAA} {Computational} {Fluid} {Dynamics} {Conference}},
    publisher = {American Institute of Aeronautics and Astronautics},
    author = {Lakshmipathy, Sunil and Togiti, Vamshi},
    month = jun,
    year = {2011},
}

@inproceedings{menter2003,
	title = {Ten Years of Industrial Experience with the SST Turbulence Model},
	booktitle = {Turbulence, heat and mass transfer},
	author = {Menter, F.R. and Kuntz, Martin and Langtry, Robin},
	volume = "4",
	pages = "625-632",
	publisher = "Begell House",
	address = "New York",
	year = "2003",
}

@article{germano_turbulence_1992,
    title = {Turbulence: the filtering approach},
    volume = {238},
    doi = {10.1017/S0022112092001733},
    journal = {Journal of Fluid Mechanics},
    author = {Germano, M.},
    year = {1992},
    pages = {325--336}
}

@misc{jakirlic_curvature_2024,
    title = {Curvature {Influence} on {Flow} and {Heat} {Transfer} in a {Concentric} {Annulus}: {Conventional} and {Sensitized} {Reynolds} {Stress} {Modeling} {Study}},
    shorttitle = {Curvature {Influence} on {Flow} and {Heat} {Transfer} in a {Concentric} {Annulus}},
    doi = {10.2139/ssrn.5061415},
    urldate = {2026-04-09},
    author = {Wang, Xiaoyu and Hussong, Jeanette and Jakirlic, Suad},
    year = {2025},
}

@article{chung_direct_2002,
    title = {Direct numerical simulation of turbulent concentric annular pipe flow: {Part} 1: {Flow} field},
    volume = {23},
    doi = {10.1016/S0142-727X(02)00140-6},
    number = {4},
    journal = {International Journal of Heat and Fluid Flow},
    author = {Chung, Seo Yoon and Rhee, Gwang Hoon and Sung, Hyung Jin},
    year = {2002},
    pages = {426--440},
}

@article{weller_tensorial_1998,
    title = {A tensorial approach to computational continuum mechanics using object-oriented techniques},
    volume = {12},
    doi = {10.1063/1.168744},
    journal = {Computers in Physics},
    author = {Weller, H. G. and Tabor, G. and Jasak, H. and Fureby, C.},
    year = {1998},
    pages = {620--631},
}

@article{issa_solution_1986,
    title = {Solution of the implicitly discretised fluid flow equations by operator-splitting},
    volume = {62},
    doi = {10.1016/0021-9991(86)90099-9},
    number = {1},
    journal = {Journal of Computational Physics},
    author = {Issa, R. I.},
    year = {1986},
    pages = {40--65},
}

@book{patankar_numerical_2018,
    edition = {1},
    title = {Numerical {Heat} {Transfer} and {Fluid} {Flow}},
    doi = {10.1201/9781482234213},
    author = {Patankar, Suhas V.},
    year = {2018},
}

@book{Wilcox,
    author = {Wilcox, David C.} ,
    title = {Turbulence Modeling for CFD.},
    publisher = {DCW Industries},
    year = {2006}
}

@article{bagheri_effects_2020,
    title = {Effects of radius ratio on turbulent concentric annular pipe flow and structures},
    volume = {86},
    issn = {0142727X},
    doi = {10.1016/j.ijheatfluidflow.2020.108725},
    urldate = {2026-04-09},
    journal = {International Journal of Heat and Fluid Flow},
    author = {Bagheri, Edris and Wang, Bing-Chen},
    month = dec,
    year = {2020},
    pages = {108725},
}

@article{banerjee_presentation_2007,
    title = {Presentation of anisotropy properties of turbulence, invariants versus eigenvalue approaches},
    volume = {8},
    doi = {10.1080/14685240701506896},
    journal = {Journal of Turbulence},
    author = {Banerjee, S. and Krahl, R. and Durst, F. and Zenger, Ch.},
    year = {2007}
}

@article{menter_scale-adaptive_2010,
    title = {The {Scale}-{Adaptive} {Simulation} {Method} for {Unsteady} {Turbulent} {Flow} {Predictions}. {Part} 1: {Theory} and {Model} {Description}},
    volume = {85},
    doi = {10.1007/s10494-010-9264-5},
    number = {1},
    journal = {Flow, Turbulence and Combustion},
    author = {Menter, F. R. and Egorov, Y.},
    year = {2010},
    pages = {113--138},
}

@article{jakirlic_extending_2015,
    title = {Extending the bounds of ‘steady’ {RANS} closures: {Toward} an instability-sensitive {Reynolds} stress model},
    volume = {51},
    issn = {0142727X},
    doi = {10.1016/j.ijheatfluidflow.2014.09.003},
    journal = {International Journal of Heat and Fluid Flow},
    author = {Jakirlić, S. and Maduta, R.},
    year = {2015},
    pages = {175--194},
}
%




\end{document}